\documentclass[a4paper,reqno]{amsart}

\date{\update}

\usepackage{amsmath,amsthm,amssymb,color,graphicx}
\usepackage{amscd,verbatim}
\usepackage[all]{xy}

\theoremstyle{plain}
\newtheorem{theorem}{Theorem}

\theoremstyle{definition}

\theoremstyle{remark}

\numberwithin{equation}{section}
\numberwithin{theorem}{section}
\numberwithin{figure}{section}
\numberwithin{table}{section}

\DeclareMathOperator{\Map}{Map}

\newcommand{\cH}{{\mathcal H}}

\newcommand{\cK}{{\mathcal K}}

\newcommand{\TT}{{\mathbb T}}
\newcommand{\ZZ}{{\mathbb Z}}

\newcommand{\sfG}{{\mathsf G}}

\newcommand{\sfN}{{\mathsf N}}
\newcommand{\PUH}{{\mathsf{PU}}(\mathcal H)}
\newcommand{\UH}{{\mathsf{U}}(\mathcal H)}
\newcommand{\sfU}{{\mathsf U}}

\newcommand{\wphi}{\widetilde{\phi}}

\begin{document}

\title[T-Duality as a Duality of Loop Group Bundles]
{T-Duality as a Duality of Loop Group Bundles}

\author{Peter Bouwknegt}
\address{
Department of Theoretical Physics,
Research School of Physics and Engineering, and 
Department of Mathematics, Mathematical Sciences Institute, 
The Australian National University, 
Canberra, ACT 0200, Australia}

\author{Varghese Mathai}
\address{Department of Pure Mathematics, University of Adelaide,
Adelaide, SA 5005, Australia}

\begin{abstract}
Representing the data of a string compactified on a circle in the 
background of H-flux in terms of the geometric data of a principal loop group bundle, 
we show that T-duality in type II string theory can be understood as the interchange 
of the momentum and winding homomorphisms of the principal loop group bundle, 
thus giving rise to a new interpretation of T-duality.
\end{abstract}
\date{}
\maketitle


\section{Introduction}
T-duality is a striking property of type II string theory on compactified spacetimes.
In earlier papers \cite{BEM}, we gave a prescription 
for the topological aspects of T-duality in type II string theory for principal
circle bundles with H-flux, and later on \cite{BHM, MR2}, 
for  higher rank principal torus bundles. In this note we 
will mainly concentrate on the case of principal circle bundles. 
In the physics 
literature, this is the same as saying that spacetime is compactified in one direction.

We begin by reviewing our results in the case of circle bundles \cite{BEM, RR88}.  Let 
\begin{equation}\label{circle1}
\xymatrix{
\TT  \ar[r] &  Y \ar[d]_{\pi}  \\
& X}
\end{equation}
be a principal circle bundle and $H$ a closed 3-form on $Y$ with integral periods.  
The T-dual of $(Y,H)$ is another spacetime $\widehat Y$ with H-flux 
$\widehat H$, with the following desirable properties:
\begin{enumerate}
\item the RR-fields and charges are mapped in a one to one manner respectively;
\item T-duality applied twice returns one to the original spacetime $Y$ with original H-flux.
\end{enumerate}
In our earlier papers, op.~cit., we generalized the Buscher rules, which are the 
local transformation rules of the low energy effective fields under T-duality 
as given in e.g.\ \cite{Buscher, Hori99},
to the the case when $H$ defines a topologically non-trivial cohomology class.
The result can be described as follows:  
$\int_{\mathbb T} H$ is a closed 2-form on $X$ with integral periods, and therefore it is 
a representative of the 1st Chern class $c_1(\widehat Y)$ of a principal circle bundle 
\begin{equation} \label{circle2}
\xymatrix{
\TT  \ar[r] &  \widehat Y \ar[d]_{\widehat\pi}  \\
& X}
\end{equation}
In fact, 
geometric prequantization asserts that $\int_{\mathbb T} H$ is the curvature of a connection
on $\widehat Y$. The T-dual flux $\widehat H$ should have the property that the closed 2-form
$\int_{\widehat{\mathbb T}} \widehat H$ with integral periods represents the 1st Chern class 
$c_1(Y)$ of the principal circle bundle $Y$. However, this does
not characterize $\widehat H$, since the lift of any closed 3-class on $X$ to $\widehat Y$ 
integrates to zero. 
To determine $\widehat H$, we need an additional condition, which is stated on 
the correspondence space $Y\times_X  \widehat Y $, defined by the commutative diagram,
\begin{equation} \label{correspondenceb}
\xymatrix @=4pc @ur { Y \ar[d]_{\pi} & 
Y\times_X  \widehat Y \ar[d]_{\widehat p} \ar[l]^{p} \\ X & \widehat Y\ar[l]^{\widehat \pi}}
\end{equation}
The additional condition that determines the cohomology class of 
the T-dual flux $\widehat H$ is that $p^*[H]= \widehat p^*[\widehat H] \in 
H^3(Y\times_X  \widehat Y, \ZZ)$.
The Gysin sequence for the circle bundle \eqref{circle2} establishes 
the existence of the T-dual flux.

One of the astonishing consequences of our theorem is that in general, there is a 
{\em change in topology} between T-dual spacetimes with T-dual background flux. 
Moreover, T-duality can be interpreted as the duality {\em exchanging the H-flux
with the Chern class}.

Using the fact that $H^3(Y,\ZZ)$ classifies (isomorphism classes) of principal
$\PUH$-bundles over $Y$, we can encode the data representing a circle bundle with 
background flux geometrically as a principal $\PUH$-bundle $\tilde\pi:P\to Y$ over a principal
$\TT$-bundle $\pi:Y\to X$.
One main goal of this note is to give a direct prescription of the T-dual spacetime 
and flux using this geometric set-up.  

The paper is organized as follows.  In Section \ref{sect:reform} we reformulate 
the geometric data in terms of certain principal loop group bundle 
along the lines of \cite{MS,BS}.  Then, in 
Section \ref{secC}, we interpret T-duality for principal circle bundles in a 
background H-flux, as a  duality of principal loop group bundles
{\em exchanging the momentum and winding homomorphisms}, and make contact 
to the classifying space analysis of \cite{BS} in Section \ref{secD}.  Our results are
summarized in Theorem \ref{thm:loop}. 

The results of this paper are in essence an amalgamation of the results of, in particular,
\cite{BV} and \cite{BS}, and based on ideas in \cite{Ev,BEM}. 
We feel, however, that it is useful to emphasize this 
geometric interpretation of T-duality as it has various other potential applications,
and in principle allows for the use of differential geometric techniques familiar to
most physicists (see, e.g., \cite{MS,Vo} for the computation of characteristic
classes of loop group bundles).
Other geometrizations of the T-duality data are possible, and have been studied in,
for example, \cite{BeHM}.

\section{Reformulation as loop group bundles}\label{sect:reform}

In this section we will review the 1--1 correspondence between 
(isomorphism classes of) principal $\sfG$-bundles $P$ over 
principal $\TT$-bundles $Y$ over $X$, and (isomorphism classes of) 
principal $L\sfG\rtimes \TT$-bundles over $X$, under the assumption
that $\sfG$ is simply connected (i.e.\ $\pi_1(\sfG)=0$) \cite{MS,BV}.
\begin{equation} \label{eqBa}
\begin{pmatrix} &&
\begin{CD}
\sfG @>>> P \\
&&  @VV\tilde\pi V \\
\TT @>>> Y \\
&& @VV\pi V \\
& & X 
\end{CD} && \end{pmatrix} \quad  \Longleftrightarrow \quad 
\begin{pmatrix} &&
\begin{CD}
L\sfG \rtimes \TT @>>> Q \\
&& @VV\Pi V \\
&& X
\end{CD} &&
\end{pmatrix}
\end{equation}
We recall the definition of $L\sfG\rtimes \TT$.   First of all, $\TT$ acts on $L\sfG$ by
$\TT \times L\sfG \to L\sfG \,,\quad (t, \gamma) \mapsto t\cdot \gamma$,
where $(t \cdot \gamma )(s) = \gamma(ts)$.
The semi-direct product $L\sfG\rtimes \TT$  is then defined
by the multiplication law
$(\gamma_1, t_1) \circ (\gamma_2, t_2) = ((t_2\cdot \gamma_1) \gamma_2, t_1t_2)$.
Equivalently, we can think of the semi-direct product  $L\sfG\rtimes \TT$
as the split short exact sequence 
\begin{equation}\label{eqBsss}
\xymatrix{
1 \ar[r] & L\sfG  \ar[r]^{\imath\ \ }  & L\sfG   \rtimes \TT  \ar[r]^{\ \ \rho} & 
\TT \ar[r] \ar@/^1.25pc/ [l]^{s} & 1 }
\end{equation}
where, for $(\gamma, t) \in L\sfG \rtimes \TT$ we have 
$\rho(\gamma,t) = t$, and $\imath(\gamma) = (\gamma,1)$.  We will refer
to $\rho$ as the `momentum homomorphism'.

We will first discuss the correspondence \eqref{eqBa} explicitly, using
transition functions. Let $\{U_i\}$ be a good cover of $X$ for which we have trivializations
$\phi_i: \pi^{-1}(U_i) \stackrel{\simeq}\to  U_i \times \TT$.  We write
$\phi_i(y) = (\pi(y) , s_i(y))$, where the `section' $s_i\,:\, \pi^{-1}(U_i) \to \TT$ 
satisfies $s_i(yt) = s_i(y) t$,
for $t\in\TT$ (group action written multiplicatively).  The transition functions
$g_{ij}\,:\,U_{ij} \to \TT$ are defined by
\begin{equation}
(\phi_i \circ \phi_j^{-1} )(x,t) = (x , g_{ij}(x) t) \,,\quad \text{i.e.}\quad
g_{ij}(x) = s_i(y) s_j(y)^{-1} \,,
\end{equation}
where $y \in \pi^{-1}(x) \subset Y$, and we write multiple 
intersections as $U_{ij}=U_i\cap U_j$, etc.  This definition does not depend
on the choice of $y \in \pi^{-1}(x)$.
The transition functions satisfy the cocycle identity,
$g_{ij}(x) g_{jk}(x) = g_{ik}(x)\,,\quad x\in U_{ijk}$.
We also recall that the $\TT$-bundle can be reconstructed from the 
transition functions by setting 
$E = \coprod_{i} ( U_i \times \TT) /\sim$, 
where we identify $(x,t)\sim (x,t')$ on $U_{ij}\times \TT$ iff $t=g_{ij}(x) t'$.

Let $V_i=\pi^{-1}(U_i)$.  Since $V_i$ is homotopic to $\TT$, and since
$\pi_1(\sfG)=0$ by assumption, the $\sfG$-bundle
$\tilde\pi:P\to Y$ trivializes over $V_i$.  Denote the local trivialization
by $\tilde \phi_i\,:\, \tilde\pi^{-1}(V_i) \stackrel{\simeq}\to V_i \times \sfG$, 
the corresponding section
by $\tilde s_i\,:\, \tilde \pi^{-1}(V_{i}) \to \sfG$, and the transition functions by 
$\tilde g_{ij}\,:\, V_{ij} \to \sfG$.

We will now show how to use these data to define a principal $L\sfG\rtimes\TT$-bundle 
$Q$ over $M$.  We define it by declaring that the transition functions 
$G_{ij} \,: \, U_{ij} \to L\sfG \times \TT$, are given by
\begin{equation}
G_{ij}(x) = (\tilde g_{ij}( \phi_j^{-1}(x,\cdot)), g_{ij}(x)) \,,\qquad x\in U_{ij}\,.
\end{equation}
Then, for $x\in U_{ijk}$, one has 
\begin{align}
G_{ij}(x) G_{jk}(x) & = (\tilde g_{ij}( \phi_j^{-1}(x,\cdot)), g_{ij}(x)) \circ
(\tilde g_{jk}( \phi_k^{-1}(x,\cdot)), g_{jk}(x)) \nonumber \\
& = (g_{jk}(x) \cdot \tilde g_{ij}( \phi_j^{-1}(x,\cdot)) \tilde g_{jk}( \phi_k^{-1}(x,\cdot)),
g_{ij}(x) g_{jk}(x)) \nonumber \\
& = (\tilde g_{ij}( \phi_k^{-1}(x,\cdot)) \tilde g_{jk}( \phi_k^{-1}(x,\cdot)),
g_{ij}(x) g_{jk}(x)) \nonumber \\
& = (\tilde g_{ik}( \phi_k^{-1}(x,\cdot)), g_{ik}(x)) = G_{ik}(x)\,,
\end{align}
where we have used the cocycle properties of the transition functions
$g_{ij}$, and $\tilde g_{ij}$.
There is an intermediate step in this construction, namely that of a $\TT$-equivariant
principal $L\sfG$-bundle $LP$ over $Y$.  This bundle is defined by its transition
functions
\begin{equation}
\tilde G_{ij}(y)(t) = \tilde g_{ij}(yt) \,,\qquad y\in V_{ij}, \ t\in\TT\,.
\end{equation}
Note that the transition functions satisfy
\begin{eqnarray}
\tilde G_{ij}(ys)(t) & = &  \tilde g_{ij}((ys)t) =  \tilde g_{ij}(y(st)) = \tilde G_{ij}(y)(st) \nonumber \\
& = & (s\cdot G_{ij}(y))(t)\,,
\end{eqnarray}
proving their equivariance.  Conversely, given a $\TT$-equivariant
principal $L\sfG$-bundle $LP$ over $Y$, we define a principal $\sfG$-bundle by
the transition functions $\tilde g_{ij}(y) = \tilde G_{ij}(y)(1)$, 
i.e.\ by evaluating the loop in $L\sfG$ at $t=1$.


\subsection{Reconstruction}
Conversely, given a
principal $L\sfG\rtimes \TT$-bundle $Q$ over $X$ with
transition functions $G_{ij}: U_{ij} \to L\sfG \rtimes \TT$, we can 
reconstruct the transition functions of a $\TT$-bundle $Y$ over $X$ and 
the $\sfG$-bundle $P$ over $Y$ as follows.  First we let 
$g_{ij} = \rho(G_{ij})$ (cf.\ \eqref{eqBsss}) be the transition function of the 
principal $\TT$-bundle $\pi : Y\to X$,
and $\tilde g_{ij} (y) = \jmath (G_{ij}(\pi(y)))(s_j(y) )$ the transition function of the 
$\sfG$-bundle, where $\jmath(\gamma,t) = \gamma$ is a left splitting of \eqref{eqBsss}.
This construction clearly is the inverse of the construction described above.
There is a more concrete way of describing the reconstruction which avoids the
explicit use of transition functions (see, e.g., \cite{MS}).
Let $Q$ be an  $L\sfG\rtimes\TT$ principal bundle.  
We define a free action of $L\sfG \rtimes \TT$ on $Q \times \sfG \times \TT$ by
$(\gamma,t) \cdot (\tilde p,g,s) = ( R_{(\gamma,t)} \tilde p, 
\gamma(s) g , st)$. 
That this defines an action of $L\sfG\rtimes\TT$ can be seen as follows
\begin{eqnarray*}
&&  (\gamma_1,t_1)  \cdot   \left(  (\gamma_2,t_2) \cdot (\tilde p,g,s) \right)   =  
(\gamma_1,t_1) \cdot ( R_{(\gamma_2,t_2)} \tilde p, \gamma_2(s) g , st_2)  \\
&& =   ( R_{(\gamma_1,t_1)} R_{(\gamma_2,t_2)} \tilde p, \gamma_1(st_2) \gamma_2(s) g , st_2),
st_1t_2) \\
&& = ( R_{(\gamma_1,t_1)\circ (\gamma_2,t_2)} \tilde p, ((\gamma_1,t_1)\circ (\gamma_2,t_2))(s),
st_1t_2)  \\
&& =  \left(  (\gamma_1,t_1)\circ (\gamma_2,t_2) \right) \cdot (\tilde p,g,s) \,.
\end{eqnarray*}
We denote by $[\tilde p,g,s]$ the equivalence class of triples under the action 
of $L\sfG\rtimes\TT$.  It is easily seen that $Q \times \sfG \times \TT$ also carries 
a free action of $\sfG$, commuting with the $L\sfG\rtimes\TT$ action, namely
\begin{equation*}
R_h (\tilde p,g,s) = (\tilde p, gh^{-1} ,s )\,,
\end{equation*}
and hence a free $\sfG$-action on $(Q \times \sfG \times \TT)/L\sfG\rtimes\TT$.
This makes $(Q \times \sfG \times \TT)/L\sfG\rtimes\TT$ into a principal $\sfG$-bundle 
over $(Q \times  \TT)/L\sfG\rtimes\TT$.  Similarly, we conclude that 
$(Q \times  \TT)/L\sfG\rtimes\TT$ is a principal $\TT$-bundle over 
$Q/L\sfG\rtimes\TT = X$.  The situation is summarized below 
\begin{equation*}
\xymatrix{
\sfG \ar[r] & (Q \times \sfG \times \TT)/ L\sfG\rtimes\TT \ar[d]_{\tilde\pi} &&  [\tilde p,g,s] \ar[d] \\
\TT \ar[r]  & (Q \times \TT) / L\sfG\rtimes\TT \ar[d]_{\pi} && [\tilde p, s] \ar[d] \\
& Q  / L\sfG\rtimes\TT && [\tilde p] }
\end{equation*}

\section{T-duality of loop group bundles}\label{secC}

In this section we will consider specifically the projective unitary group 
$\sfG=\PUH$ on a separable Hilbert space $\cH$, which is 
the geometric set-up for the discussion of T-duality for $\TT$-bundles 
with H-flux, as discussed in the introduction.  In this case we will
see that there is an additional circle `hidden' in the $L\sfG\rtimes \TT$-bundle
description and that by interchanging the role of the circles
(interchanging momentum and winding) before applying the reconstruction 
process produces the T-dual bundles.  Schematically
\begin{equation*}
\begin{pmatrix} &&
\begin{CD}
\sfG @>>> P \\
&&  @VV\tilde\pi V \\
\TT @>>> Y \\
&& @VV\pi V \\
& & X 
\end{CD} && \end{pmatrix} 
 \Rightarrow 
\begin{pmatrix} &&
\begin{CD}
L\sfG \rtimes \TT @>>> Q \\
&& @VV\Pi V \\
&& X
\end{CD} &&
\end{pmatrix}
  \Rightarrow 
\begin{pmatrix} &&
\begin{CD}
\sfG @>>> \widehat P \\
&&  @VV\widehat{\tilde\pi} V \\
\TT @>>> \widehat Y \\
&& @VV\widehat \pi V \\
& & X 
\end{CD} && \end{pmatrix} 
\end{equation*}

First recall that the loop group $L\sfG = \Map(\TT,\sfG)$, and the 
based loop group $\Omega\sfG = \{ \gamma\in L\sfG \ | \ \gamma(0)=e\}$, where
$e\in\sfG$ is the identity element, are related by the 
isomorphism of groups 
$L\sfG \cong \Omega \sfG \rtimes \sfG$, 
$\gamma (\cdot) \mapsto (\gamma(\cdot) \gamma(0)^{-1} , \gamma(0) )$, 
where the action of $\sfG$ on $\Omega\sfG$ is given by
$(g \cdot \gamma)(\cdot) = g \gamma(\cdot) g^{-1}$.
Now observe that 
\begin{eqnarray*}
&& \pi_1( L\sfG \rtimes \TT)  \cong \pi_1(L\sfG) \oplus \pi_1(\TT)   \cong
\pi_1( \Omega \sfG \rtimes \sfG)  \oplus \pi_1(\TT) \\
&&  \cong 
\pi_1(\sfG) \oplus \pi_1(\Omega \sfG) \oplus \pi_1 (\TT)  \cong
\pi_1(\sfG) \oplus \pi_2(\sfG) \oplus \pi_1(\TT)\,.
\end{eqnarray*}
Now, if $\sfG$ is a simply connected (as we have assumed 
previously), compact Lie group, then $\pi_1(\sfG) \cong \pi_2(\sfG)\cong 0$ 
and hence $\pi_1( L\sfG \rtimes \TT) \cong \ZZ$, i.e.\ the group
$L\sfG \rtimes \TT$ `contains' only one circle described by Eqn.~\eqref{eqBsss}.
If, however, $\sfG=\PUH$, then $\pi_2(\sfG) \cong \ZZ$ and thus 
$\pi_1( L\sfG \rtimes \TT) \cong \ZZ \oplus \ZZ$ which means we have two
circles sitting inside $L\sfG \rtimes \TT$.

The second circle can be recovered as follows.
The group $\sfG=\PUH$ has a canonical central extension $\UH$
\begin{equation} \label{eqCb}
\xymatrix{ 1 \ar[r] &  \TT \ar[r] &  \UH \ar[r] &  \PUH \ar[r]& 1 }
\end{equation}
The central extension \eqref{eqCb} has a connection compatible with the group 
structure.  The holonomy of this connection is a homomorphism
$\text{hol} \colon  L\sfG \to \TT$.
We let $\sfN = \text{Ker}( \text{hol} \, : \, L\sfG \to \TT)$, a normal subgroup of 
$L\sfG$.  We thus have an exact sequence
\begin{equation}\label{eqBsst}
\xymatrix{
1 \ar[r] & \sfN \rtimes \TT  \ar[r]^{\imath\ \ }  & L\sfG   \rtimes \TT  \ar[r]^{\ \ \omega} & 
\TT \ar[r]  & 1 
}
\end{equation}
where $\omega$, the winding homomorphism, is defined by $\omega(\gamma, t) = 
\text{hol}(\gamma)$.
If this were again a split exact sequence, like \eqref{eqBsss}, we could apply
the reconstruction theorem and obtain a dual $\sfG$-bundle over a dual $\TT$-bundle.
However, the sequence \eqref{eqBsst} is not split and hence $\sfN\rtimes \TT$ is not 
isomorphic to $L\sfG$.  From the exact sequence in homotopy it follows easily,
however, that $\sfN \rtimes \TT$ and $L\sfG$ are homotopy equivalent.
After performing this homotopy, we can perform the reconstruction.  The result being
our T-dual circle bundle with T-dual H-flux described in the introduction.  A proof 
of this statement follows from the work of Bunke and Schick \cite{BS}.  The connection of 
the above discussion to this work is the subject of the next section.

\section{Classifying maps} \label{secD}

Let us reformulate the equivalence \eqref{eqBa} in terms of classifying maps,
specialize to the case $\sfG=\PUH$, and make the connection to the work
of \cite{BS}.

The principal $\TT$-bundle $\pi:Y\to X$ gives us a map $\wphi: X\to B\TT$, while
the the principal $\sfG$-bundle $\tilde \pi: P\to Y$ gives us a map $\psi: Y \to B\sfG$.  
We have already seen that the principal $\sfG$-bundle $\tilde \pi: P\to Y$ is equivalent to 
a $\TT$-equivariant principal $L\sfG$ bundle $\widehat \pi : LP\to Y$, hence a 
classifying map $\widehat\psi: Y\to BL\sfG$ satisfying 
$\widehat \psi(yt) = t\cdot \widehat\psi(y)$.  Also, we
can construct a map $\widehat\phi : Y \to E\TT$, covering $\wphi$.  To summarize
\begin{equation*}
\begin{matrix}
\begin{CD}
Y @>\widehat\phi>> E\TT \\
@V\pi VV  @VV\widehat\pi V \\
X @>\wphi>> B\TT
\end{CD}
\end{matrix} \qquad\qquad
\begin{matrix}
\begin{CD}
Y @>\widehat\psi>> BL\sfG \\
@V\psi VV   \\
B\sfG 
\end{CD}
\end{matrix}
\end{equation*}
The two maps $\widehat\phi$ and $\widehat\psi$ can be combined into a map
$\Phi = (\widehat\phi, \widehat\psi) : Y \to E\TT \times  BL\sfG$, which descends to a map $\phi:X \to
E\TT \times_\TT  BL\sfG$, since
$\phi(yt) = [ \widehat \phi (yt) , \widehat \psi (yt)] = [ \widehat \phi (y) t , t\cdot \widehat\psi(y)] 
= [ \widehat \phi (y)  ,  \widehat\psi(y)]  = \phi(y)$, 
where $[\ , \ ]$ denotes equivalence classes of pairs in $E\TT \times  BL\sfG$ 
under the action of $\TT$.  Let us denote $R = E\TT \times_\TT  BL\sfG$.

Conversely, suppose we are given 
a map $\phi:X \to R$.  
Since $\TT$ acts freely on $E\TT$ we have a projection 
$\widehat\pi: R \to B\TT$.  I.e.\ $R$ is a principal $BL\sfG$ bundle over $B\TT$.
\begin{equation*}
\xymatrix{
BL\sfG \ar[r] &  R \ar[d]^{\widehat\pi} \\
& B\TT }
\end{equation*}
Thus we have a map $\tilde\phi = \widehat\pi \circ\phi: X \to B\TT$.
This map defines our principal $\TT$-bundle $\pi: Y\to X$.  We can lift $\wphi$ to
$\widehat\phi:Y \to E\TT$, and then define a map $\widehat \psi : Y \to BL\sfG$ uniquely by 
$\phi( \pi(y) ) = [ \widehat\phi (y) , \widehat\psi(y) ]$, and we conclude that $\widehat\psi$ is 
$\TT$-equivariant. Finally, we obtain a map $\psi: Y \to B\sfG$, defining
our pricipal $\sfG$-bundle over $Y$, by using 
that $BL\sfG \cong B(\Omega\sfG \rtimes \sfG) \cong
E\sfG \times _\sfG B\Omega\sfG$, so that in particular we have a map $\widehat\pi :
BL\sfG \to B\sfG$.    To summarize, we have shown that 
$R \cong B(L\sfG \rtimes \TT)$, i.e.\ the map $\phi$ classifies
principal $L\sfG \rtimes \TT$-bundles over $X$.
The various maps are summarized in the diagram below
\begin{equation*}
\xymatrix@=3pc{
&& \TT \ar[d] & \TT \ar[d] \\
B\sfG   & BL\sfG \ar[l]^{\widehat\pi}   \ar[d] & Y \ar@/_1.5pc/ [ll]  \ar[l]^\psi \ar[r]_{\widehat \phi}  
\ar[d] &  E\TT \ar[d] \\
& B(L\sfG \rtimes \TT) \ar@/_1.5pc/ [rr]^{\widehat \pi} &  X \ar[l]_\phi \ar[r]^{\tilde\phi}  & B\TT 
}
\end{equation*}
\bigskip

Now we specialize to $\sfG = \PUH$.  We recall that $\TT=  K(\ZZ,1)$ and 
$\sfG =K(\ZZ,2)$, hence $B\TT = K(\ZZ,2)$, and $B\sfG=K(\ZZ,3)$.
We have seen that $B(L\sfG \rtimes \TT)$, the classifying space for principal 
$L\sfG \rtimes \TT$-bundles over $X$ is isomorphic 
to $R= E\TT \times_\TT BL\sfG$, and that $R$ therefore has the natural structure
of a principal $BL\sfG \cong (K(\ZZ,3) \rtimes K(\ZZ,2))$-bundle over $B\TT=K(\ZZ,2)$.
However, there is another way of interpreting $R$, namely as 
a $K(\ZZ, 3)$ homotopy fibration over $K(\ZZ, 2) \times K(\ZZ, 2)$ \cite{BS} 
(see also \cite{MR2}). 
Moreover, 
there is a map $T\colon R \to R$ such that $T^*\colon H^2(R, \ZZ) \to H^2(R, \ZZ)$ 
exchanges the two generators, and 
$T\circ T$ is homotopic to the identity on $R$.  It turns out that $T\colon R \to R$ 
implements T-duality for principal circle bundles with background flux (see \cite{BS}):
A principal $L\sfG \rtimes \TT$-bundle $Q$ over $X$ has two natural characteristic classes 
of degree 2 on $X$. One of these
is the first Chern class of the associated circle bundle over $X$, $c_1(Y)$, and the 
other is given by integration over the fibre of $Y$, of the Dixmier-Douady invariant of $P$. 
We denote these by $c(Q)$ and $d(Q)$ respectively, and they are the pullback under 
the classifying map $\phi:X \to R$ of the generators of $ H^2(R, \ZZ)$. 

Hence, in terms of classifying maps, the 
the T-dual principal 
$L\sfG \rtimes \TT$-bundle $\widehat Q$ over $X$ is defined by considering the
continuous map $T \circ f : X \to B(L\sfG \rtimes \TT)$, and by associating to it 
$\widehat Q =(T \circ f)^*(E(L\sfG \rtimes \TT))$.  It follows that 
T-duality exchanges the entries of the pair $(c(Q),d(Q))$, and that T-duality 
applied twice 
gives a bundle that is isomorphic to $Q$, since $T\circ T \sim {\rm I}_R$.
We summarize this as follows.

\begin{theorem}[T-duality as a duality of principal loop group bundles]\label{thm:loop}
Given a principal loop group bundle,
$$
\begin{CD}
L\sfG \rtimes \TT @>>> Q \\
&& @VV\Pi V \\
&& X
\end{CD} 
$$
with classifying map $\phi :X \to B(L\sfG \rtimes \TT)$, then there exists a T-dual 
loop group principal bundle,
$$
\begin{CD}
L\sfG \rtimes \TT @>>>\widehat  Q \\
&& @VV\widehat \Pi V \\
&& X
\end{CD} 
$$
with classifying map $T\circ \phi:X \to B(L\sfG \rtimes \TT)$, which has the following
properties:
\begin{enumerate}
\item  $\widehat{\widehat  Q}$ is isomorphic to $Q$;
\item  $c(\widehat Q) = d(Q)$ and $d(\widehat Q) = c(Q)$.
\end{enumerate}
Geometrically speaking, T-duality can be viewed as the exchange,
of the momentum and winding homomorphisms of the previous section 
(see Eqns.\ \eqref{eqBsss} and \eqref{eqBsst}).
\end{theorem}

\section{Concluding remarks}

The RR-fields in type IIA string theory can also be interpreted as principal loop 
group bundles over $X$.  To this end, recall that in
Theorem 7.2(5) of \cite{BCMMS}, the following description of twisted K-theory 
is given. Every element of twisted K-theory, twisted by a principal $\sfG$-bundle 
$P$ over $Y$ (here as before, $\sfG=\PUH$), is given by a principal $\sfU_\cK \rtimes \sfG$
bundle $A$ over $Y$ that projects onto $P$. Here $\sfU_\cK$ denotes the group of 
unitary operators in the given Hilbert space which are  
of the form  identity plus compact operator (known as the universal gauge group \cite{HM}), 
and the projection  condition means that 
$P \cong A\times_{\sfU_\cK \rtimes \sfG} \sfG$. 
Therefore RR-fields can be interpreted as
principal $L(U_\cK \rtimes \sfG)\rtimes \TT$-bundles over $X$ that project onto the 
given principal $L(\sfG)\rtimes \TT$-bundle $Q$ over $X$.
Similarly, charges of type IIA  RR-fields can 
be interpreted as a principal $\Omega\sfU_\cK \rtimes \sfG$
bundles $B$ over $Y$ that project onto $Q$. We conclude that the T-duality isomorphism of the 
RR-fields and their charges in type II string theory as established in
in \cite{BEM}, can be also interpreted as a duality isomorphism between loop group 
bundles on $X$. 

Finally, the analysis of this paper can be generalized to T-duality for higher rank
torus bundles, provided the principal $\sfG$-bundle $P$ is trivial when restricted to
the torus fibre of $Y\to X$.  This requires the H-flux to be classical, in the terminology of
\cite{BHM}, and replaces the condition that $\pi_1(\sfG)=0$.\bigskip

\noindent {\bf Acknowledgments:} This research was supported under the 
Australian Research Council's Discovery Projects funding scheme.


\end{document}